\documentclass[11.5pt, titlepage]{article}
\usepackage[a4paper,margin=3cm]{geometry}
\usepackage{graphicx}
\usepackage{tabularx}
\usepackage{amssymb}
\usepackage{booktabs}	
\usepackage{url}
\usepackage{tocloft}
\usepackage{lineno}
\usepackage{amsfonts, amsmath, microtype, parskip, url}
\usepackage{graphicx, xcolor,mathpazo}
\usepackage{array}
\usepackage{color, colortbl}
\usepackage[noperiod]{jabbrv}
\usepackage[normalem]{ulem}

\definecolor{OliveGreen}{rgb}{0,0.6,0}
\usepackage{marginnote}
\newcommand{\points}[1]{
	\label{#1}
	\leavevmode\marginnote{\makebox[\marginparwidth][r]
		{\normalsize \color{OliveGreen} #1}}\ignorespaces}

\usepackage{xparse}
\DeclareDocumentCommand \myedit { o m } {%
	\IfNoValueF {#1} {\points{#1}}
	{#2}
}

\newcommand{\mydel}[1]{}

\begin{document}
\Large{\textbf{Burst detection methods}	}
\\\Large{Ellese Cotterill, Stephen J. Eglen\\
Department of Applied Mathematics and Theoretical Physics\\
University of Cambridge\\
CB3 0WA UK}
\bigskip
\\\textbf{Abstract}
\\`Bursting', defined as periods of high frequency firing of a neuron separated by periods of quiescence, has been observed in various neuronal systems, both \textit{in vitro} and \textit{in vivo}. It has been associated with a range of neuronal processes, including efficient information transfer and the formation of functional networks during development, and has been shown to be sensitive to genetic and pharmacological manipulations. Accurate detection of periods of bursting activity is thus an important aspect of characterising both spontaneous and evoked neuronal network activity. A wide variety of computational methods have been developed to detect periods of bursting in spike trains recorded from neuronal networks. In this chapter, we review several of the most popular and successful of these methods.

\subsection*{Table of Abbreviations}
\begin{tabular}{ll}
  \hline
  CMA & Cumulative Moving Average\\
  IQR & Inter-Quartile Range\\
  IRT & ISI Rank Threshold\\
  ISI & InterSpike Interval\\
  LTD & Long Term Depression\\
  LTP & Long Term Potentiation\\
  MEA & MultiElectrode Array\\
  MI & Max Interval\\
  PS & Poisson Surprise\\
  RS & Rank Surprise\\
  RGS & Robust Gaussian Surprise\\
  \hline
\end{tabular}
\section{Introduction}
\begin{figure}[h]
	\centering
	\includegraphics[width=140mm]{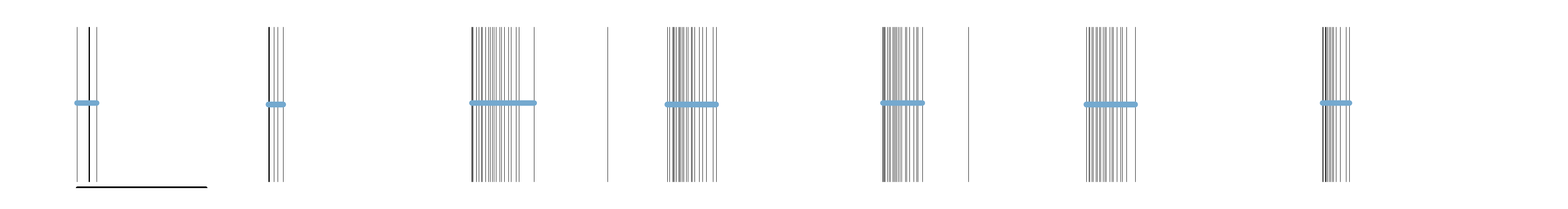}
	\caption[Example of bursting activity recorded in a mouse retinal ganglion cell]{Example of bursting activity in a spike train recorded from mouse retinal ganglion cells. Horizontal blue lines show the location of bursts. Scale bar represents 1$\,$s.}
	\label{burst_eg}
\end{figure}

\begin{figure}[h]
	\centering \vspace{3mm}
	\includegraphics[width=110mm]{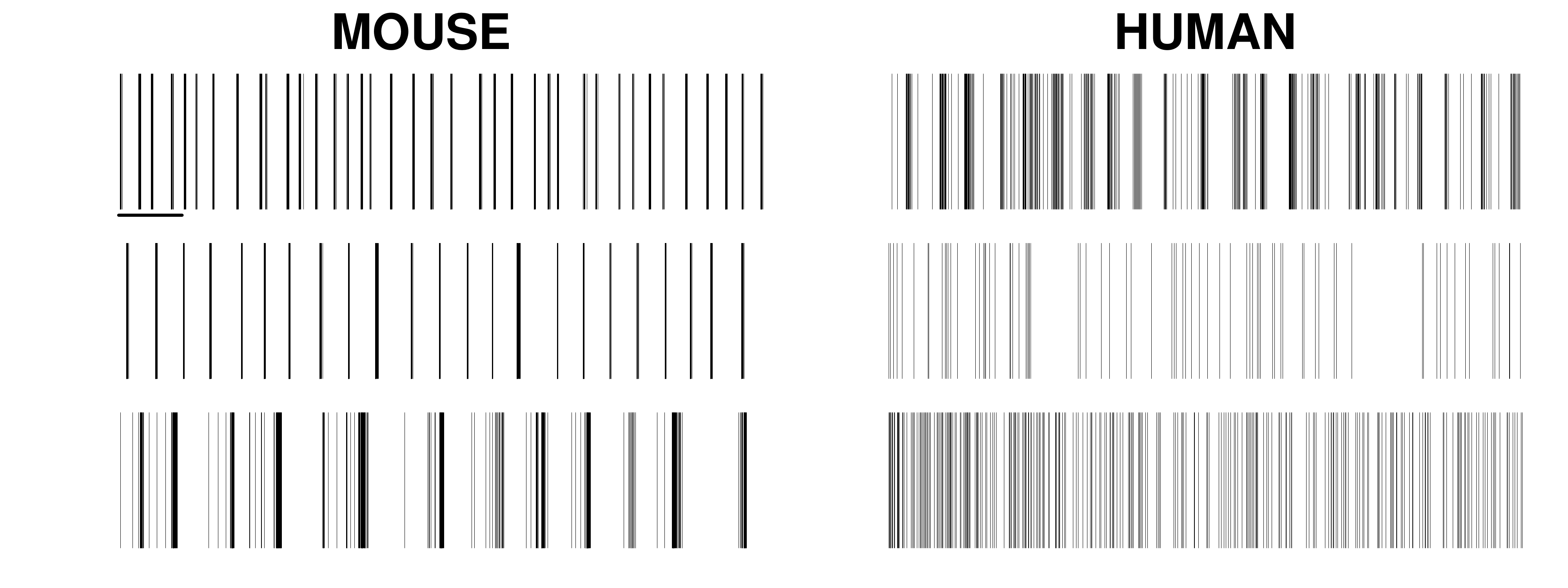}
	\caption[Examples of spike trains recorded from mouse and human neuronal networks]{Examples of spike trains from mouse and human neuronal networks. Each row represents the spikes recorded from one electrode and the scale bar represents 30$\,$s. Recordings from human neuronal networks often exhibit more variable and complex spontaneous activity patterns.}
	\label{mouse_hum_egs}
\end{figure}

Neuronal bursting, observed as intermittent periods of elevated spiking rate of a neuron (see Figure \ref{burst_eg}), has been observed extensively in both \textit{in vitro} and \textit{in vivo} neuronal networks across various network types and species \cite{Weyand2001, Chiappalone2005, Pasquale2010}. \myedit[]{These bursts can be isolated to a single neuron or, commonly, occur simultaneously across many neurons, in the form of `network bursts' \cite{VanPelt2004a, Wagenaar2006, Pasquale2008, Bakkum2013}}
\\\\Bursting activity is believed to play a role in a range of physiological processes, including synapse formation \cite{Maeda1995} and long-term potentiation \cite{Lisman1997}. Analysis of patterns of bursting activity can thus be used as a proxy for studying the underlying physiological processes and structural features of neuronal networks. A common method of studying bursting activity \textit{in vitro} involves the use of MEA recordings of spontaneous or evoked neuronal network activity \cite{Lonardoni2015, Charlesworth2015, Pimashkin2011, VanPelt2004a}. This approach has been employed to study changes in spontaneous network activity over development \cite{Wagenaar2006}, and the effect of pharmacological or genetic manipulations \cite{Eisenman2015, Charlesworth2016}.
\\ \\ Despite the importance of bursting and its prevalence as a feature used to analyse neuronal network activity, there remains a lack of agreement in the field about the definitive formal definition of a burst \cite{Cocatre-Zilgien1992, Gourevitch2007}. There is also no single technique that has been widely adopted for identifying the location of bursts in spike trains. Instead, a large variety of burst detection methods have been proposed, many of which have been developed and assessed using specific data sets and single experimental conditions. As most studies of bursting activity have been performed on experimental data from recordings of rodent \myedit[]{\mydel{or feline}} neuronal networks \cite{Charlesworth2015, Mazzoni2007}, this type of data has most often been used to assess the performance of burst detection techniques \cite{Chiappalone2005, Mazzoni2007, Gourevitch2007}. 
\\ \\Recently, it has been shown that networks of neurons derived from human stem cells can be grown successfully on MEAs and exhibit spontaneous electrical activity, including bursting \cite{Illes2007, Heikkila2009}. Human stem-cell derived neuronal cultures have also been demonstrated to be a suitable alternative to rodent neuronal networks in applications such as neurotoxicity testing \cite{Yla-Outinen2010}. This has \myedit[]{\mydel{lead} led}to a demand for a robust method of analysing bursting in these networks, which commonly exhibit more variable and complex patterns of bursting activity than rodent neuronal networks \cite{Kapucu2012} (see Figure \ref{mouse_hum_egs}). Recently, some burst detection methods have been developed which specifically focus on analysing bursting activity in these types of variable networks \cite{Kapucu2012, Valkki2017}.
\myedit[]{\section{Physiological significance of neuronal bursting}}
\myedit{Neuronal bursting is a frequently observed phenomenon in MEA recordings of cultures of dissociated neurons, as well as in numerous \textit{in vitro} systems \cite{Wagenaar2006, Pasquale2008, Weyand2001, Legendy1985}. In cultured rodent cortical networks, bursts, and in particular synchronised `network bursts' generally arise as a feature of the spontaneous network activity after around 1 week \textit{in vitro} \cite{Kamioka1996}. Most studies observe that these network bursts then increase in frequency and size before reaching a peak around 3 weeks \textit{in vitro} \cite{VanPelt2004b, VanPelt2004a, Chiappalone2006}. This peak in network bursting activity generally corresponds to the period in which the synaptic density of the network reaches its maximum \cite{VanHuisen1985, Kamioka1996, VanPelt2004b}. This is followed by a period of shortening of network bursts, which coincides with a stage of 'pruning' or reduction in dendritic spine synapses and maturation of excitatory connections between neurons \cite{Chiappalone2006, Illes2007, Ichikawa1993, VanPelt2005}.  As well as being correlated with neuronal network development and maturation, bursting patterns of spontaneous activity are also believed to play an important role in regulating cell survival. High frequency bursting has been shown to increase neuronal survival in cortical cultures, while suppression of spontaneous activity has been observed to greatly increase rates of programmed cell death \cite{Golbs2011, Heck2008}.  
\\\\Bursting has also been observed to be involved with a range of physiological processes in mature neuronal networks. For example, bursting is believed to be a more efficient method of information transfer between neurons than single spikes. Central synapses in various brain regions have been shown to exhibit low probabilities of neurotransmitter release in response to single presynaptic spikes, making information transfer by single spikes unreliable \cite{Borst2010, Branco2009, Allen1994}. However, bursts of spikes can lead to `facilitation', a process in which a rapid succession of spikes leads to a build up of intracellular Ca$^{2+}$ in the presynaptic terminal. This increases the probability of neurotransmitter release and resultant production of EPSPs with subsequent spikes \cite{Thomson1997, Krahe2004}. In addition to being involved in these mechanisms of short term plasticity, bursting has also been implicated in long term potentiation (LTP) and depression (LTD). For example, in the hippocampus, postsynaptic bursting at temporally relevant intervals could produce long term synaptic changes \cite{Pike2004, Froemke2006,Thomas1998}. 
\\\\It has also been suggested that bursts of spikes transmit information with a higher signal-to-noise ratio than single spikes \cite{Sherman2001}. Evidence of this has been seen in a variety of brain regions, such as the hippocampus, where place-fields have been shown to be more accurately defined by bursts than individual spikes \cite{Otto1991}. Bursting has also been shown to produce sharper sensory tuning curves \cite{Cattaneo1981, Krahe2004} and more reliable feature extraction than single spikes \cite{Gabbiani1996, Sherman2001, Krahe2002}.
\\\\The importance of neuronal bursting has also been demonstrated through its association with a variety of behaviours \textit{in vivo}, including visual processing, reward and goal directed behaviour and sleep and resting conditions \cite{Cattaneo1981, Krahe2004, Tobler2003, Schultz1997, Schultz1998, Evarts1964, Barrionuevo1981, McCarley1983, Weyand2001, Steriade2001}. Bursting of hippocampal place cells has also been observed during exploration of new environments \cite{OKeefe1993, Epsztein2011}. The presence of bursting in these, as well as other memory-related behaviours \cite{Burgos-Robles2007, Xu2012}, suggests that bursting plays a specific role in memory and learning in the adult brain \cite{Paulsen2000}.
\\\\Additionally, bursting activity has been seen to be altered in certain pathological conditions \cite{Walker2008, Jackson2004, Miller2011, Singh2016}. For example, increased bursting activity has been observed in the basal ganglia of Parkinson’s patients, with correlations between the level of bursting activity and the progression of the disease \cite{Lobb2014, Ni2001}. This suggests that the study of bursting activity could not only reveal important features of normal brain function, but also how this is altered in diseased states.}
\\\\\myedit[]{\mydel{Role of bursting in neuronal networks}}
\myedit[]{*deleted this section*}
	\section{Previous approaches to burst detection}
Since the development of the first methods to identify bursting in neuronal networks more than three decades ago, many techniques have been proposed. These methods take a variety of approaches. 
\myedit[]{\subsection{Fixed threshold-based methods}} 
The simplest approaches involve imposing thresholds on values such as the minimum firing rate or maximum allowed interspike interval (ISI) within a burst, and classifying any sequence of consecutive spikes satisfying these thresholds as a burst.  In well-ordered spike trains, these thresholds can be set as fixed values by visual inspection \cite{Weyand2001, Chiappalone2005}. Other methods also incorporate additional thresholds on relevant parameters such as the minimum interval between two bursts and the minimum duration of a burst, to restrict detected bursts to those with biologically realistic properties \cite{NexTechnologies2014}.
\myedit[]{\subsection{Adaptive threshold-based methods}} 
As opposed to having fixed threshold parameters that are chosen by the user, other burst detection algorithms derive the values of their threshold parameters adaptively from properties of the data, such as the mean ISI \cite{Chen2009} or total spiking rate \cite{Pimashkin2011}. Commonly, this involves the use of some form of the distribution of ISIs on a spike train. For spike trains containing bursting activity, the smoothed histogram of ISIs on the train should have a peak in the region of short ISIs, which represents within-burst ISIs, and one or more peaks at higher ISI values, representing intra-burst intervals. A threshold for the maximum ISI allowed within a burst can be set at the ISI value representing the turning point in the histogram \cite{Cocatre-Zilgien1992}. 
\\ \\Several other adaptive burst detection algorithms also use distributions related to the ISI histogram to calculate thresholds for burst detection. \cite{Selinger2007} and \cite{Pasquale2010} argue that the histogram of log(ISI)s provides a better separation of within and between-burst intervals, and use this histogram to set the threshold for the maximum within-burst ISI at the minimum between the first two well separated peaks. \cite{Kaneoke1996} use the histogram of discharge density rather than ISIs for burst detection, \myedit[]{\mydel{while \cite{Bakkum2013} employ the ISI histogram between every $n$th spike in a network,}} while \cite{Kapucu2012} derive the threshold parameters for detecting bursts in their algorithm from the cumulative moving average of the ISI histogram.
\myedit[]{\subsection{Surprised-based methods}} 
Another category of burst detection techniques are the surprise-based methods, which use statistical techniques to distinguish periods of bursting from baseline neuronal firing. The earliest of such methods was developed by \cite{Legendy1985}, and detects bursts as periods of deviation from an assumed underlying Poisson process of neuronal firing. This method critically assumes Poisson-distributed spike trains, which has been shown to be inappropriate for many common spike trains, in particular because of the refractory period between spikes \cite{Cateau2006}. Despite this, the Poisson Surprise method has been one of the most widely used burst detection methods since its development over thirty years ago \myedit[]{(398 citations as of June 2018)} and is still commonly used for analysing bursting activity in experimental studies of numerous neuronal network types \cite{Singh2016, Pluta2015, Senn2014}. More recently, other surprise-based burst detectors have been developed that replace the assumption that baseline firing follows a Poisson process with other assumptions about the underlying distribution of spikes \cite{Ko2012, Gourevitch2007}.
\myedit[]{\subsection{Other methods}} 
Other burst detectors take alternative approaches to separate bursting from background spiking activity. \cite{Turnbull2005} examine the slope of the plot of spike time against spike number to detect bursts as periods of high instantaneous slope. \cite{Martinson1997} require bursts to be separated by intervals at least two standard deviations greater than their average within-burst ISIs, while \cite{Tam2002} propose a parameter-free burst detection method, in which sequences of spikes are classified as bursts if the sum of their within-bursts ISIs is less than the ISIs immediately before and after the burst. 
\\ \\ Numerous studies have also used various forms of Hidden Markov Models to analyse neuronal activity patterns \cite{Radons1994, Chen2009a, Abeles1995}. These methods assume that a neuron stochastically alternates between two or more states, characterised by differences in their levels of activity. \cite{Tokdar2010} apply this idea to burst detection by modelling neuronal activity using hidden semi-Markov models.
\subsection{Burst detection methods}
\begin{figure}[h]
	\centering
	\includegraphics[width=150mm]{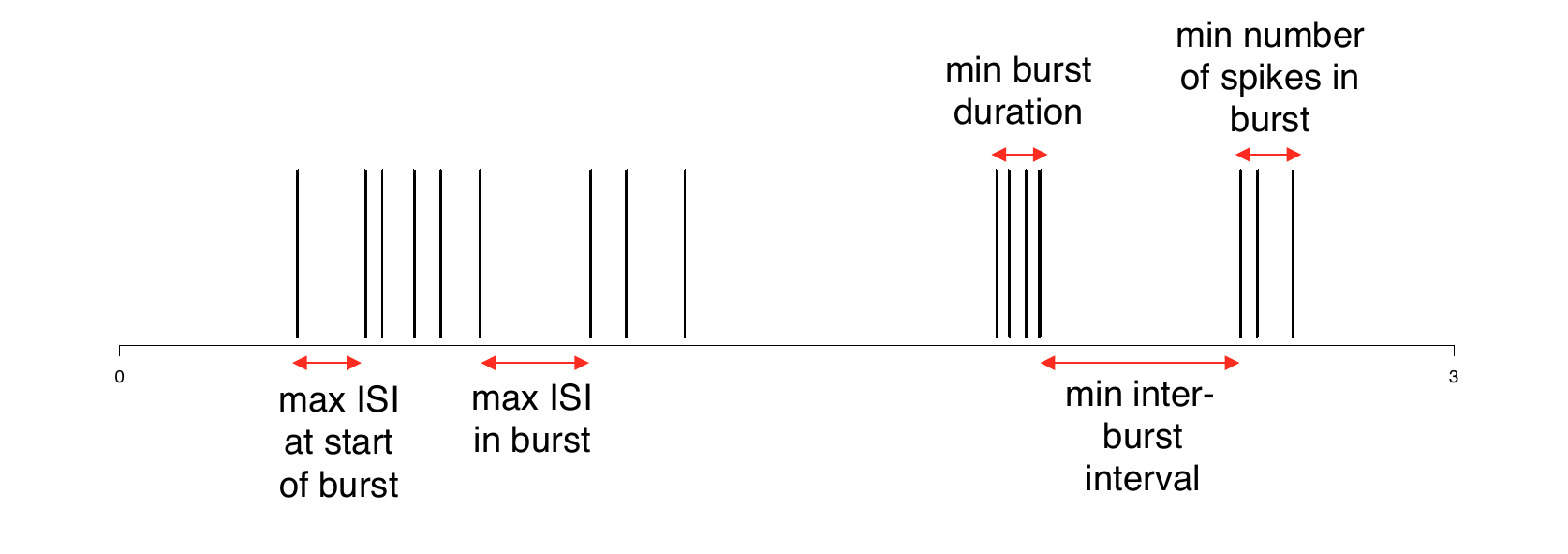}
	\caption[Illustration of the parameters used by the MaxInterval method]{Illustration of the parameters used by the MaxInterval method.}
	\label{mi_eg}
\end{figure}

\begin{figure}[h]
	\centering
	\includegraphics[width=150mm]{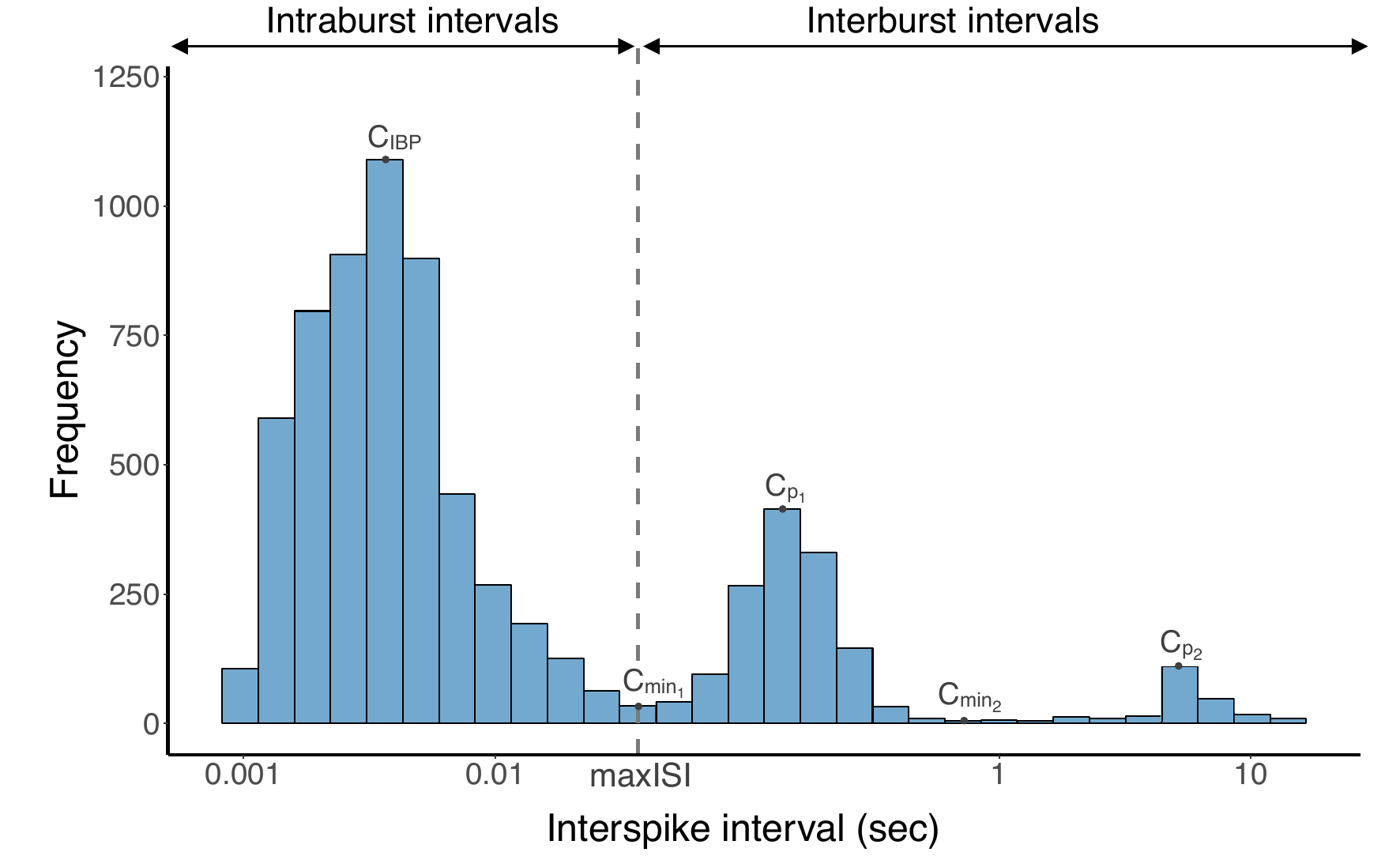}
	\caption[Example of log-adjusted ISI histogram with the threshold for intraburst and interburst intervals found using the logISI method]{Example of log-adjusted ISI histogram with the threshold for intraburst and interburst intervals found using the logISI method.}
	\label{logISI_eg}
\end{figure}

\begin{figure}[h]
	\centering
	\includegraphics[width=145mm]{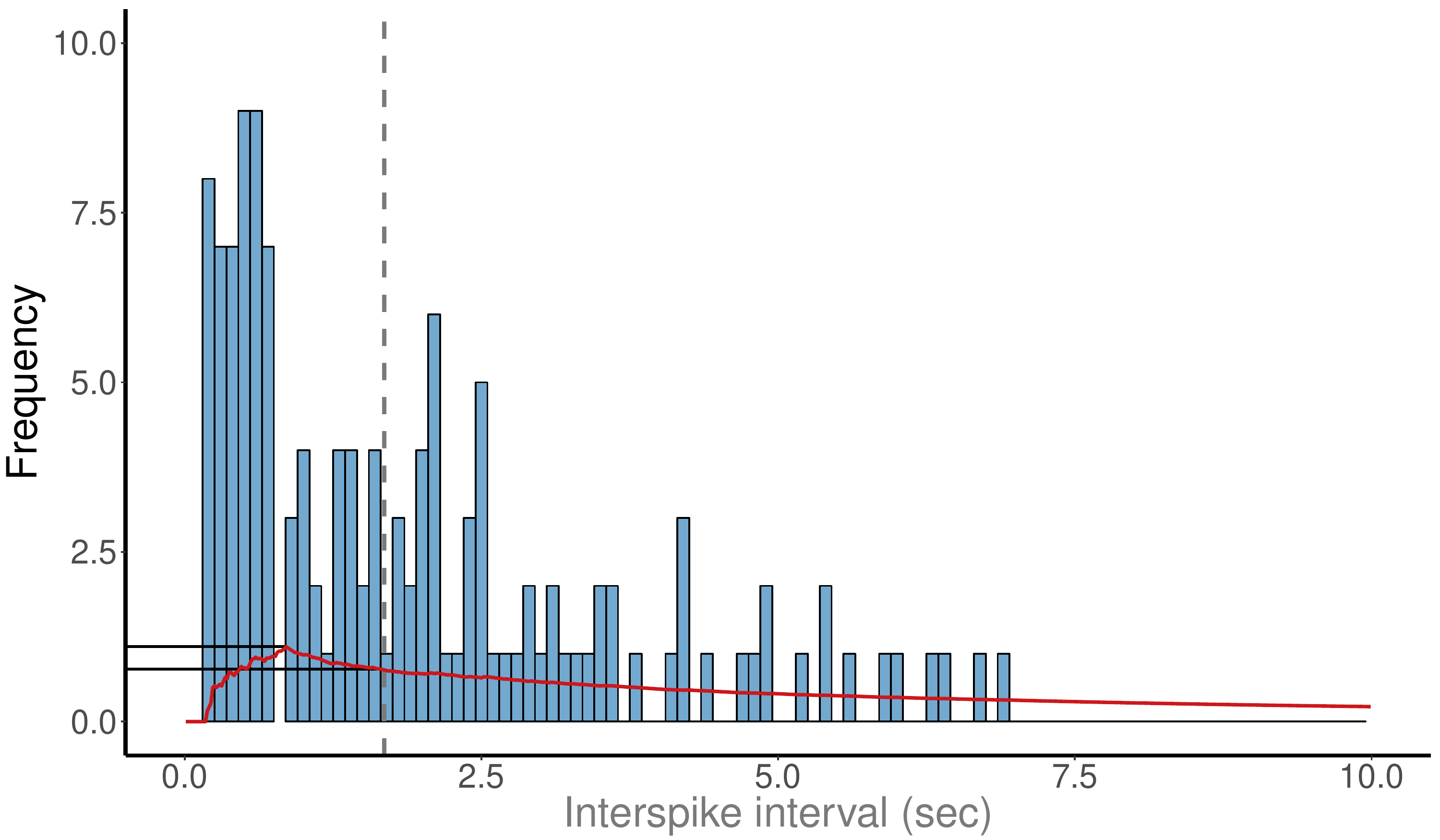}
	\caption[Example of ISI histogram with the threshold for intraburst and interburst intervals found using the CMA method]{Example of ISI histogram with the threshold for intraburst and interburst intervals found using the CMA method. Red line shows the cumulative moving average of the ISI histogram.}
	\label{CMA_eg}
\end{figure}

In this section, we will outline a number of key existing burst detection algorithms. Given the vast number of available burst detection techniques, the following have been chosen for their relevance and popularity in the existing literature\myedit[]{, and represent examples of each of the approaches to burst detection outlined above.} 
\begin{table}[h]
	\centering
	\begin{tabular}{lp{6cm}c}
          \toprule \textbf{Abbrev.}& \textbf{Method} & \textbf{Reference} 
		\\ \midrule \multicolumn{3}{c}{\textbf{Fixed threshold-based methods}}
		\\ \textbf{MI} & MaxInterval  & \cite{NexTechnologies2014} 
		\\ \multicolumn{3}{c}{\textbf{Adaptive threshold-based methods}}
		\\ \textbf{logISI} &    LogISI &  \cite{Pasquale2010} 
		\\ \textbf{CMA} & Cumulative Moving Average & \cite{Kapucu2012}
		\\\textbf{IRT} & ISI Rank Threshold 
		& \cite{Hennig2011} 
		\\ \multicolumn{3}{c}{\textbf{Surprise-based methods}}
		\\ \textbf{PS} &  Poisson Surprise & \cite{Legendy1985}
		\\  \textbf{RS} & Rank Surprise&  \cite{Gourevitch2007}
		\\  \textbf{RGS} & Robust Gaussian Surprise & \cite{Ko2012} 
		\\ \multicolumn{3}{c}{\textbf{Other methods}}
		\\ \textbf{HSMM} & Hidden Semi-Markov Model  & \cite{Tokdar2010}
		\\ \bottomrule
	\end{tabular}
	\caption[]{\myedit{Burst detectors classified by their approach to burst detection.}  }
	\label{burst_param_vals}
\end{table}
\\ \\ \textit{MaxInterval method \cite{NexTechnologies2014}}
\\ Bursts are defined using five fixed threshold parameters, shown in Figure \ref{mi_eg}. The value of these parameters are chosen a priori and any series of spikes that satisfy these thresholds is classified as a burst.
\\\\\textit{LogISI method \cite{Pasquale2010}}
\\The histogram of log(ISI)s on a spike train is computed, using a bin size of 0.1 in $\log(\text{ISI})$ units. Let $C_k$ denote the ISI count in the $k$th bin of this histogram, which corresponds to an ISI size of ISI$_k$, and MCV denote a pre-specified threshold value, known as the maximum cutoff value. The location of the peaks of this histogram are found using a custom peak finding algorithm described in \cite{Pasquale2010}. The largest peak of the histogram corresponding to an ISI less than or equal to MCV is set as the intraburst peak, $C_{IBP}$. If no peak is found in the histogram with ISI$_k \leq$ MCV, the spike train is classified as containing no bursts.
\\ \\ In the case that an intraburst peak is present, the minimum value of the histogram between the intraburst peak and each of the following peaks, $C_{p_i}$  ($i=1, ..., N$), is found. For each minimum, a void parameter is calculated that represents how well the corresponding peak is separated from the intraburst peak, as
\[ void(i) = 1-\frac{C_{min_i}}{\sqrt{C_{IBP}\cdot C_{p_i}}}\]
where $C_{min_i}$ is the minimum value of $C_k$ for $IBP<k<p_i$. 
\\\\The smallest ISI$_{min_i}$ for which $void(i)>0.7$ is set as the threshold for the maximum ISI in a burst, $maxISI$ (see Figure \ref{logISI_eg}). Any series of at least three spikes separated by ISIs less than $maxISI$ are classified as bursts. If no point with a void value above 0.7 is found, or if $maxISI>$ MCV, bursts are detected using MCV as the threshold for the maximum ISI in a burst and then extended to include spikes within $maxISI$ of the beginning or end of each of these bursts.
\vspace{1mm}
\\\\\textit{Cumulative Moving Average (CMA) method \cite{Kapucu2012}}
\\ This method also uses the histogram of ISIs on a spike train. The cumulative moving average (CMA) at each ISI bin of the histogram is calculated. The CMA of the $N$th ISI bin is defined as 
\[CMA_N=\frac{1}{N}\sum^N_{k=1}C_k\, ,\] where $C_k$ is the ISI count in the $k$th bin. The skewness of the CMA distribution is used to determine the values of two threshold parameters, $\alpha_1$ and $\alpha_2$, based on the scale given in \cite{Kapucu2012}. The maximum of the CMA distribution, $CMA_{max}$, is found and the value of $maxISI$ is set at the ISI bin at which the CMA is closest in value to $\alpha_1 \cdot CMA_{max}$ (see Figure \ref{CMA_eg}). Burst cores are then found as any sequences of at least three spikes separated by ISIs less than $maxISI$.
\\ \\ \cite{Kapucu2012} suggest extending these burst cores to include burst-related spikes. These are found using a second cutoff, set at the value of the ISI bin at which the CMA is closest to $\alpha_2\cdot CMA_{max}$. Spikes within this cutoff distance from the beginning or end of the existing burst cores are classified as burst-related spikes. For this study, only the burst cores detected by this method were examined, omitting any burst related spikes.
\\\\\textit{ISI rank threshold method \cite{Hennig2011}} 
\\ In the ISI rank threshold (IRT) method, the rank of each ISI on a spike train relative to the largest ISI on the train is calculated, with $R(t)$ denoting the rank of the ISI beginning at time $t$. The probability distribution, $P(C)$, of spike counts in one-second time bins over the spike train is also found. A rank threshold, $\theta_R$, is set to a fixed value, and a spike count threshold, $\theta_C$, is calculated from $P(C)$. A burst is then defined to begin at a spike at time $t$ if the rank of the proceeding ISI satisfies $R(t)<\theta_R$ and the spike count in the following second, $C(t, t+1)$, exceeds $\theta_C$. The burst continues until a spike is found for which $C(t, t+1)<\frac{\theta_C}{2}$. 
\\\\\textit{Poisson Surprise method \cite{Legendy1985}}
\\ The average firing rate, $\lambda$, on a spike train is calculated, and the underlying activity on this spike train is assumed to follow a Poisson process with rate $\lambda$. The Poisson Surprise (PS) statistic for any period of length $T$ containing $N$ spikes is calculated as 
\[S=-\log P\]
where 
\[P = \exp\left(-\lambda T \sum_{n=N}^\infty \frac{(\lambda T)^n}{n!}\right)\]
is the probability that $N$ or more spikes occur randomly in a period of length $T$. 
\\ \\A surprise maximization algorithm described in \cite{Legendy1985} is then used to find the set of bursts that maximises the PS statistic across the entire spike train. This involves initially identifying bursts as any sequence of three consecutive spikes separated by ISIs which are less than half of the mean ISI on the spike train. Spikes are then added to the end and removed from the beginning of each of these initial bursts until the sequence of spikes with the maximum PS statistic is found. Finally, any bursts which have a PS statistic below a pre-defined threshold level are discarded.  
\\\\ \textit{Rank Surprise method \cite{Gourevitch2007}}
\\ The Rank Surprise (RS) burst detection algorithm is a non-parametric adaptation of the Poisson Surprise approach. To implement this method, all ISIs on a spike train are ranked by size, with the smallest ISI given a rank of one. In the absence of any bursting activity, the ISI ranks should by independently and uniformly distributed. For any period containing $N$ spikes separated by $N-1$ ISIs with ranks $r_n$, ..., $r_{n+N-1}$, the Rank Surprise statistic is defined as 
\[RS=-\log(P(D_N\leq r_n+ ... + r_{n+N-1}))\]
where $D_N$ is the discrete uniform sum distribution between 1 and $N$ and $r_n$ is the rank of the $n$th ISI on the spike train.
\\ \\Bursts are then chosen to maximise the RS statistic across the entire spike train using an exhaustive surprise maximisation algorithm, outlined in \cite{Gourevitch2007}. A fixed threshold for $maxISI$ is first calculated from the distribution of ISIs on the spike train. The first sequence of at least three spikes with ISIs less than $maxISI$ are found and an exhaustive search of all of the subsequences of ISIs within this period is performed to find the subsequence with the highest RS value. If this value is above a fixed minimum significance threshold, chosen a priori, it is labelled as a burst. This process is repeated on the remaining ISI subsequences within the period of interest until all significant bursts are found. Following this, the next sequence of spikes with ISIs below $maxISI$ is examined in a similar fashion, and this process is continued until the end of the spike train.
\\ \\ \textit{Robust Gaussian Surprise method \cite{Ko2012}}
\\ In the Robust Gaussian Surprise (RGS) method, the distribution of log(ISI)s on each spike train is found and centred around zero. The normalised log(ISI)s from each spike train in the study are then pooled and the central distribution of this joint data set is found using a procedure outlined in \cite{Ko2012}. A burst detection threshold for $maxISI$ is set at the 0.5 percentile of this central distribution, which is estimated as 2.58 times the median absolute deviation of the distribution.
\\ \\The Gaussian burst surprise value in any interval on a spike train is defined as 
\[GS_B=-\log(P)\] where $P$ is the probability that the sum of normalised log(ISI)s in the interval is greater than or equal to the sum of an equal number of i.i.d. Gaussian random variables with mean and variance equal to that of the central distribution. 
\\ \\ Any consecutive sequence of spikes separated by intervals less than $maxISI$ are classified as burst cores. These burst cores are then extended by adding intervals to the beginning and end of the burst cores until the sequence with the maximum value of $GS_B$ is found. In the case of overlapping bursts, the burst with the largest $GS_B$ value is retained. Finally, any detected bursts with $GS_B$ below a predefined threshold value are discarded. \cite{Ko2012} also propose a similar method for identifying pauses in spike trains.
\\ \\\textit{Hidden Semi-Markov Model method \cite{Tokdar2010}}
\\This method is based on the assumption that neurons switch stochastically between two states: `non-bursting' (state 0) and `bursting' (state 1), that can be modelled using a Hidden semi-Markov model. The transition times between the two states are modelled using two Gamma distributions, $f_0^{ITI}$ and $f_1^{ITI}$. Within each of the states, the ISI times are modelled using two additional gamma distributions, $f_0^{ISI}$ and $f_1^{ISI}$. The parameters of these four distributions are learned from the data. A custom Markov chain Monte-Carlo algorithm described in \cite{Tokdar2010} is then used to compute the posterior probability that a neuron is in a bursting state at any given time.  A fixed threshold value is chosen a priori, and any periods during which the posterior probability exceeds this value are classified as bursts. 
\subsection{Evaluation of burst detection techniques}
\begin{figure}[h]
	\centering \vspace{5mm}
	\includegraphics[width=145mm]{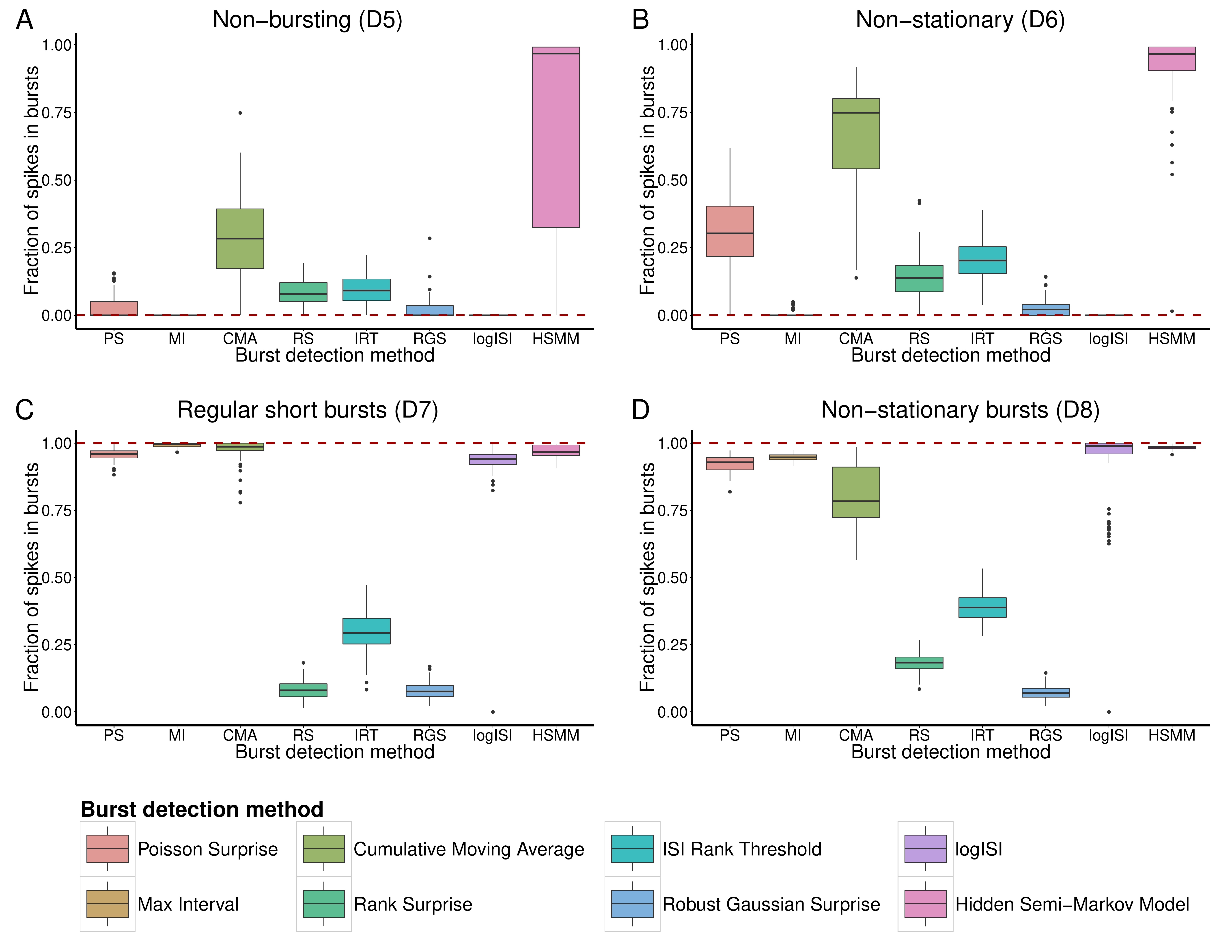}
	\caption{Fraction of spikes in bursts found by each burst
          detector in 100 synthetic trains with \textbf{A} no bursting
          (D5), \textbf{B}  no bursting and non-stationary firing rate
          (D6), \textbf{C} short regular bursts (D7), \textbf{D}
          bursts with non-stationary burst lengths and durations
          (D8). Dotted line shows desired result from an ideal burst
          detector; methods close to this line are deemed to work
          well. In each `box-and-whisker' plot, boxes show the median
          $\pm$ inter-quartile range (IQR), and whiskers extend to
          median $\pm \ 1.5\times$ IQR. Outliers are represented as
          points. Figure reproduced with permission from  \protect\cite{Cotterill2016c}.}
	\label{sim_results1}
\end{figure}

In \cite{Cotterill2016c}, we performed a thorough evaluation of the burst detection methods outlined above. This involved first assessing the methods against a list of desirable properties that we deemed an ideal burst detector should possess \myedit[]{(see Table \ref{des_prop}). This was achieved by generating synthetic spike trains with specific properties of interest to represent each desirable property. The output of each burst detector when used to analyse each set of spike trains was then compared to the `ground truth' bursting activity.} 
	\begin{table}[h] \centering 	
	\begin{tabular}{lm{12.5cm}}
		\toprule\multicolumn{2}{l}{\textbf{Desirable properties}}
		\\ \midrule
		D1 & Deterministic: The method should detect the same bursts over repeated runs on the same  data, to ensure consistency and reproducibility of results
		\\ \midrule D2 & No assumption of spike train distribution: The method should not assume ISIs follow a standard statistical  
		distribution, to ensure wide applicability to a variety of spike trains
		\\ \midrule  D3 & Number of parameters: The method should have few parameters, to reduce the variability inherently introduced through parameter choice
		\\\midrule D4 & Computational time: The method should run in a reasonable amount of time using standard personal computers
		\\ \midrule D5 & Non-bursting trains: The method should detect few spikes as being within bursts in spike trains containing no obvious bursting behaviour
		\\ \midrule D6 & Non-stationary trains: The method should detect few spikes as being within bursts in spike trains with non-stationary firing rates that contain no obvious bursting behaviour 
		\\ \midrule D7 & Regular short bursts: The method should detect a high proportion of spikes in bursts in spike trains containing short well-separated bursts
		\\  \midrule D8 & Non-stationary bursts: The method should detect a high proportion of spikes in bursts in spike trains containing bursts with variable durations and numbers of spikes per burst
		\\ \midrule D8 & Regular long bursts: The method should detect a high proportion of spikes in bursts and accurate  number of bursts in spike trains containing long bursts with low within-burst firing rates
		\\ \midrule   D10& High frequency bursts: The method should detect a high proportion of spikes in bursts and accurate number of bursts in spike trains containing a large number of short bursts
		\\ \midrule   D11 & Noisy train: The method should classify a high number of within-burst spikes as bursting and a low number of interburst spikes as bursting in spike trains containing both bursts and noise spikes
		\\ \bottomrule
	\end{tabular} \caption{\myedit[]{Desirable properties for a burst detector. Table reproduced from \protect\cite{Cotterill2016c}.}}
	\label{des_prop}
\end{table}\myedit[]{\mydel{For example, an ideal burst detector should identify small amounts of bursting activity in spike trains simulated to contain no bursting activity, and accurately detect bursting periods in spike trains simulated to contain various types of bursting activity, including regular short and long bursts, and bursts with non-stationary properties (see Figure \ref{sim_results1} for example results).} Figure \ref{sim_results1} shows the performance of the chosen burst detectors on a sample of these properties. Most burst detectors can accurately detect a small amount of bursting activity in spike trains simulated to contain no bursting behaviour (Figure \ref{sim_results1}A), with the exception of the HSMM and CMA methods, which detect a significant amount of erroneous bursting. Conversely, most burst detectors accurately identified most  bursting activity in spike trains containing only regular short bursts (Figure \ref{sim_results1}C). However, the RS, IRT and RGS methods performed poorly here, only detecting a small proportion of the bursting activity.}
\\\\This approach \myedit[]{of assessing the performance of each burst detection method against desirable properties} allowed us to determine a ranking for each of the burst detectors, in which the Rank Surprise, Robust Gaussian Surprise and ISI Rank Threshold methods ranked particularly poorly \myedit[]{(see Table \ref{des_results})}. 
		\begin{table} \centering  \begin{tabular}{lcccccccc}
		\toprule
		\textbf{Desirable property} & \multicolumn{8}{c}{\textbf{Burst detection method}} 
		\\  \cmidrule(r{4pt}){1-1} \cmidrule{2-9}& PS & MI & CMA & RS & IRT & RGS & logISI & HSMM
		\\ \cmidrule{2-9}
		D1 Deterministic & \checkmark & \checkmark & \checkmark & \checkmark & \checkmark & \checkmark & \checkmark & $\times$
		\\ D2  Distribution assumption&  $\times$ & \checkmark & \checkmark & \checkmark & \checkmark & $\times$ & \checkmark & $\times$
		\\  D3  Number of parameters&  \checkmark & $\times$ & \checkmark & \checkmark & \checkmark & \checkmark & \checkmark & $\times$
		\\D4  Computational time & \checkmark & \checkmark & \checkmark & \checkmark & \checkmark & \checkmark & \checkmark & $\times$
		\\ \cmidrule{2-9}  D5 Non-bursting & 4 & 1 & 7 & 5 & 6 & 3 & 1 &  8
		\\D6 Non-stationary & 6 & 2 & 7 & 4 & 5 & 3 & 1 & 8 
		\\ D7 Regular bursting & 4 & 1 & 2 & 7 & 6 & 7 & 5 & 3
		\\D8 Non-stat bursts & 4 & 3 & 5 & 7 & 6 & 8 & 2 & 1
		\\  D9 Long bursts & 2 & 4 & 3 & 8 & 5 & 7 & 6 & 1
		\\ D10 High frequency & 5 & 1 & 4 & 7 & 6 & 8 & 2 & 3
		\\  D11 Noisy bursts & 5 & 1 & 2 & 7 & 6 & 8 & 4 & 2
		\\ \midrule  Total (Relative rank) & 30 (4) & 13 (1) &  30 (4) & 45 (8) & 40 (6) & 44 (7) & 21 (2) & 26 (3)
		\\ \bottomrule
	\end{tabular}
	\caption[]{\myedit[]{The performance of each method on the desirable properties specified in Table \ref{des_prop}. For binary properties, D1–-D4, each method was judged to either possess the property or not, while for properties D5–-D11, the performance of each method was ranked against the other methods (1=best, 8=worst) and summed to produce an overall ranking. Table adapted from \protect\cite{Cotterill2016c}.}} \label{des_results}
\end{table}	
Further assessment of the burst detectors was then achieved by examining the coherence of the bursts detected by each method with visually annotated bursts in experimental recordings of mouse retinal ganglion cells (RGCs). This allowed us to analyse the specificity and sensitivity of the burst detectors as their input parameters were varied. This analysis reinforced the low levels of adaptability of the RS, RGS and IRT methods at analysing this type of data. The HSMM method was also seen to have a consistently high false positive rate compared to other burst detectors used to analyse this data.
\\\\Based on these assessments, four burst detectors, namely the MI, logISI, PS and CMA methods, were chosen as the best performing burst detection methods, and used to analyse bursting activity in novel recordings of networks of human induced pluripotent stem cell (hiPSC)-derived neuronal networks over several months of development. This analysis showed a slight increase in the proportion of bursting activity observed in these networks as they mature, although this increase was far lower than that which has been observed in developing rodent neuronal networks \cite{Charlesworth2015, Chiappalone2005, Wagenaar2006}.
\\\\From this analysis, we concluded that no existing burst detector possesses all of the desirable properties required for `perfect' identification of bursting periods in highly variable networks. The CMA and PS methods possessed many of the desirable properties, but had limitations such as their tendency to overestimate bursting activity in spike trains containing sparse or no bursting activity, particularly those with a non-stationary firing rate. 
\\\\ Overall, the MI and logISI methods showed the most promise for
achieving robust burst analysis in a range of contexts. These methods
possessed most properties we deemed desirable for a burst detection
method and were generally able to achieve high coherence with visually
detected bursts in experimental MEA recordings. These methods,
however, still had limitations. The MI method requires the choice of
five parameters, the optimal values of which can be challenging to
determine, particularly when analysing recordings from a variety of
experimental conditions \cite{Cotterill2016c}. The logISI method had a tendency to underestimate bursting in some spike trains, particularly those with non-standard bursting activity.
\\\\The overall recommendation from this analysis was to choose a burst detector from the several high performing methods outlined above based on the number of freedom the user wishes to control. The MI method is a good first choice for these purposes, and despite the large number of parameters this method requires, these parameters are easy to interpret biologically and adjust to achieve the desired burst detection results for the specific situations in which it is utilised. If appropriate parameters cannot be found for the MI method, a high performing alternative is the logISI method, which can be implemented without choosing any input parameters. This method is most effective when there is a clear distinction between the size of within-burst and between-burst intervals on a spike train. In cases when this distinction is not apparent, the PS and CMA methods are reasonably effective alternative burst detection methods, however post hoc screening for outliers in terms of burst duration is advisable when using either of these methods.
\\\\One robust approach to burst detection would be to use several burst detectors to analyse the data of interest, and compare the results of each method. If the burst detectors are largely in agreement, this provides confidence in the nature of the bursting activity identified in the experimental data. Any major discrepancies between the results from the methods can also be used to identify areas where one or more burst detectors may be performing poorly, which can be further investigated through inspection of the specific spike trains of interest.
\subsection{Network-wide burst detection}
As well as single neuron bursts, synchronous bursting of networks of neurons, termed `network bursts', are a ubiquitous feature of various neuronal networks. In rat cortical cultures, these network bursts have been observed to arise from around one week \textit{in vitro}, and comprise the dominant form of spontaneous network activity at this age \cite{Chiappalone2005, VanPelt2004b}. Network bursts increase in frequency and size before reaching a peak at around 3 weeks \textit{in vitro}, corresponding to the period in which synaptic density in the network reaches is maximum \cite{VanPelt2004b, VanPelt2004a, Chiappalone2006}. 
\\\\As well as in rat cortical cultures, the presence of network bursting activity has also been observed in a variety of other brain regions and species \textit{in vitro} \cite{VanDenPol1996, Ben-Ari2001, Rhoades1994, Harris2002, Meister1991} and \textit{in vivo} \cite{Chiu2001, Leinekugel2002, Weliky1999}. Recently, synchronous bursting resembling that in rat cortical cultures has also been observed in networks produced from human embryonic or induced pluripotent stem-cell derived neurons, generally arising 8 to 12 weeks after differentiation and increasing in frequency over development \cite{Heikkila2009, Odawara2016, Amin2016}. 
\subsubsection{Existing network burst detection techniques}
A variety of techniques have been developed to detect these network-wide bursts. Several of these methods identify bursts as increases in the network-wide firing rate \cite{Mazzoni2007, Raichman2008}. These periods, however, do not necessarily consist of single neurons bursts across multiple electrodes. Other methods define network bursts only when single-neuron bursts occur simultaneously across numerous recorded electrodes \cite{Wagenaar2006, Pasquale2010}. \myedit[]{For example, \cite{Bakkum2013} combine the spikes detected on all channels of an MEA into a single spike train and employ the ISI histogram between every $n$th spike in this network-wide spike train to determine an appropriate threshold for the maximum ISI within a network burst. \cite{Wagenaar2005}, on the other hand, detect `burstlets' on each electrode individually using an adaptive threshold based on the electrode's average firing rate. A network burst is then defined as any period in which burstlets on multiple electrodes overlap.}
\\\\Network-wide information can also be incorporated into single-neuron burst detection techniques to improve their performance.  \cite{Martens2014} showed that the peaks corresponding to intra and interburst spikes in an ISI histogram were better separated when pooled ISIs from multiple electrodes of an MEA were included, rather than simply those from a single spike train. They also proposed a pre-processing technique designed to improve the detection of bursts, particularly on noisy data. This involves creating a return map, which plots the ISI immediately proceeding each spike ($ISI_{pre}$) against the ISI following the spike ($ISI_{post}$). Background spikes lie in the region of this graph with both high $ISI_{pre}$ and $ISI_{post}$, and are removed from consideration by the burst detection method. The performance of various single channel burst detection techniques were shown to be significantly improved when applied to data pre-processed in this way, compared to the original data \cite{Martens2014}. 
\\\\ Additionally, \cite{Valkki2017} adapted the CMA method of \cite{Kapucu2012} to incorporate information from multiple MEA electrodes. In this multi-CMA method, instead of individual histograms for each spike train, the ISI histogram from the combined ISIs from multiple electrodes are used to calculate the threshold for burst detection in an identical method to the original CMA method. This threshold is then used to detect bursts on each electrode individually. The electrodes that are used for combined analysis by this method can be chosen from a variety of options, including analysing all electrodes in a single MEA simultaneously, analysing the spike trains from a single electrode over several experimental time points, or analysing all electrodes over all time points in the experiment. This adaptation has been shown to reduce the number of excessively long sparse bursts identified by the original CMA method, improving its performance at analysing highly variable spike trains.
\subsection{Summary and future directions}
In this article we have summarised the main techniques of burst
detection.  Moving from an informal definition ("bursts are groups of
spikes that are close to each other in time") to a formal mathematical
definition has proved challenging.  Our experience is that when the
datasets are relatively clean, there is good agreement between methods.
However, when the data are noisy, not only \myedit[]{\mydel{to} do} different methods \myedit[]{\mydel{agree} disagree},
different human observers will also disagree.  Here we have outlined
several of the methods that we believe work relatively well, but are
fallible when presented with noisy data.  Future work in this area might
be centred around developing methods that are more robust to noisy
data. \myedit[]{Possible steps towards this may involve generating more realistic synthetic datasets to train and assess burst detection techniques, or the incorporation of noise-reducing preprocessing steps prior to burst detection, such as those developed by \cite{Martens2014}} \\\\\ \myedit[]{\mydel{Furthermore, we have yet to rigorously survey other fields of
time series analysis to see what other methods for detecting clustering 
of event times are available.} Outside of neuroscience, the detection of `bursty' events is also a more general problem in time series analysis. For example, identifying
bursts of gamma rays can aid in the detection of black holes, and the  detection of periods of high trading volume of a stock is of relevance to regulators looking for insider trading \cite{Zhu2003}. Various techniques have been developed for detecting bursting periods in these and other data types, including sliding window and infinite state automaton-based models \cite{Zhu2003, Zhang2006, Kleinberg2002, Boyack2004, Kumar2003}.  Ideas from these burst detectors developed in other domains may be
useful for informing future approaches to burst detection in a neuroscience context. 
\\\\The increasing use of high density MEAs, which contain up to several thousand electrodes \cite{Maccione2014, Lonardoni2015}, to record \textit{in vitro} neuronal activity as well as the prevalence of multi-well MEAs in
applications such as high-throughput neurotoxicity screening \cite{Valdivia2014, Nicolas2014} and drug safety testing \cite{Gilchrist2015} also has implications for burst detection. In particular, the computational complexity of burst detection methods becomes increasingly relevant in these high-throughput situations, as does the importance of minimising the manual intervention required to run the burst detectors, such as through autonomous parameter selection. The development of online burst detection techniques that can detect bursting activity in real time is also necessary to facilitate areas such as the study of real time learning in embodied cultured networks, and applications involving bidirectional communication between
biological tissue and computer interfaces \cite{Wagenaar2005, Bakkum2004}. This is another area in which ideas adopted from burst detectors developed outside of neuroscience may benefit the field. 
\\\\In conclusion, years of study of bursting activity in cultured neuronal networks has led to the development of many promising burst detection methods. However, a `perfect' method for analysing bursting activity remains elusive. In the future, the development of improved burst detection methods will be essential to keep up with advances in experimental techniques used to record bursting activity, such as the use of higher density arrays and availability of recordings from human stem cell-derived networks.}

\subsection{Acknowledgements}
EC was supported by a Wellcome Trust PhD Studentship and a National Institute for Health Research (NIHR) Cambridge Biomedical Research Centre Studentship.

\subsection{Other resources}
\begin{itemize}
	\item Open source R code for the burst detection methods
          outlined in this chapter are available at
          \url{https://github.com/ellesec/burstanalysis} and archived
          at \url{https://doi.org/10.5281/zenodo.1284064}.
\end{itemize}
\bibliographystyle{abbrv} 
\bibliography{Bursting_chapter}
\end{document}